\documentclass[groupedaddress,
 reprint,
 amsmath,amssymb,
 aps,
 prd,
 superscriptaddress,
 nofootinbib
]{revtex4-1}

\usepackage{aas_macros_new}
\usepackage{graphicx}
\usepackage{dcolumn}
\usepackage{bm}
\usepackage{hyperref}
\usepackage[utf8]{inputenc}
\usepackage{braket}
\usepackage{dsfont}
\usepackage{color}
\usepackage[dvipsnames]{xcolor}
\usepackage{soul}
\usepackage{flushend}

\newcommand{\ldb}{\lambda_\mathrm{dB}}

\begin{document}

\title{{\sc AxioNyx}: Simulating Mixed Fuzzy and Cold Dark Matter}

\author{Bodo Schwabe}
\email{bschwabe@astro.physik.uni-goettingen.de}
\affiliation{
 Institut f\"ur Astrophysik, Universit\"at G\"ottingen, Germany
}

\author{Mateja Gosenca}
\email{mateja.gosenca@auckland.ac.nz}
\affiliation{Department of Physics, University of Auckland, New Zealand}

\author{Christoph Behrens}
\email{christoph.behrens@uni-goettingen.de}
\affiliation{
 Institut f\"ur Astrophysik, Universit\"at G\"ottingen, Germany
}

\author{Jens C. Niemeyer}
\email{jens.niemeyer@phys.uni-goettingen.de}
\affiliation{
 Institut f\"ur Astrophysik, Universit\"at G\"ottingen, Germany
}
\affiliation{Department of Physics, University of Auckland, New Zealand}

\author{Richard Easther}
\email{r.easther@auckland.ac.nz}
\affiliation{Department of Physics, University of Auckland, New Zealand}

\date{\today}

\begin{abstract}
The distinctive effects of fuzzy dark matter are most visible at non-linear galactic scales. We present the first simulations of mixed fuzzy and cold dark matter, obtained with an extended version of the {\sc Nyx} code. Fuzzy (or ultralight, or axion-like) dark matter dynamics are governed by the comoving Schr\"odinger-Poisson equation.  This is evolved with a pseudospectral algorithm on the root grid, and with finite differencing at up to six levels of adaptive  refinement. Cold dark matter is evolved with the existing N-body implementation in {\sc Nyx}. We present the first investigations of spherical collapse in mixed dark matter models, focusing on radial density profiles, velocity spectra and soliton formation in collapsed halos. We find that the effective granule masses decrease in proportion to the fraction of fuzzy dark matter which quadratically suppresses soliton growth, and that a central soliton only forms if the fuzzy dark matter fraction is greater than 10\%. The {\sc Nyx} framework supports baryonic physics and key astrophysical processes such as star formation. Consequently, {\sc AxioNyx} will enable increasingly realistic studies of fuzzy dark matter astrophysics. 
\end{abstract}

\maketitle
\section{Introduction}
\label{sec:intro} 

The physical nature of dark matter is a major open question for both astrophysics and particle  physics. Axion-like particles (ALPs) are viable candidates \cite{Turner1983, Khlopov1985,Press1990,Hu2000, Sin_1994, Sahni2000,Matos2000,Guzman2000,Goodman_2000, Peebles_2000,Amendola_2006,Hwang2009}; see Refs~\cite{Marsh2016a,NIEMEYER2020} for recent reviews.   
Among them, ultralight axions predict the strongest differences from cold dark matter (CDM) on galactic scales. String theory motivates the existence of large numbers of scalar fields arising from the compactification of extra dimensions with inherently low masses and weak interactions \cite{Arvanitaki2010,Hui2017}. Scenarios with more than a single species of dark matter, or mixed dark matter (MDM) models, arise naturally from this perspective.

Independent of their fundamental origin, extremely light scalar particles with negligible interactions produced in a coherent non-thermal state via the misalignment mechanism are typically referred to as Fuzzy Dark Matter (FDM) \cite{Hu2000} or Ultralight Dark Matter (ULDM)~\cite{Sin_1994}. Unlike QCD axions, their masses can be low enough to exhibit wave-like behaviour on astrophysical scales, up to distances on the order of a kiloparsec given current  bounds on the axion mass. Chaotic interference structures with strong, short-lived density variations and halos with gravitationally-bound central solitons are among the potentially observable wave-like phenomena \cite{Schive2014}. These structures have no direct analogues in CDM.  However,  on scales greater than the de Broglie wavelength, $\ldb=h/mv$, the properties of collapsed structures are indistinguishable from those of pressureless CDM, via the Schrödinger-Vlasov correspondence \cite{Widrow1993UsingMatter,Uhlemann2014}. 

Small-scale differences open the possibility of observational comparisons of FDM and CDM. Moreover, the observed small scale properties of galaxies may be in tension with simple CDM  scenarios, and FDM appears to  ameliorate some of these issues \cite{Marsh2014,Schive2014,Marsh2014AxionProblem,Gonzales2017,Bernal2018,Kendall2019}. On the other hand, FDM solitons alone may not explain the observed radial density profiles of galactic cores \cite{Deng2018,Bar2018,Robles2019,Burkert2020}. 
However, these scales are also heavily influenced by baryonic physics, star formation, supernovae, environmental effects and galaxy-galaxy interactions, all of which operate in the collapsed, nonlinear regime. Consequently, obtaining accurate predictions for FDM dynamics in realistic astrophysical environments is a challenging task even for state-of-the-art simulations on modern supercomputers.
 
FDM is composed of non-relativistic  scalar matter described by a wavefunction $\psi$, which both sources and interacts with the  Newtonian gravitational potential. Consequently, FDM is governed by the coupled Schr\"odinger-Poisson equation. The local FDM velocity is represented by the gradient of the phase of the complex-valued wavefunction. Extracting this from simulations requires that their spatial resolution is fine enough to resolve $\ldb$, not only in small high-density regions, but also in
extensive
low density regions if high-speed flows are present. This makes full cosmological simulations with FDM
significantly more challenging than their CDM analogues, which can be efficiently evolved in phase space with N-body algorithms. 

Several different approaches for solving the Schr\"odinger-Poisson equations numerically have been employed in the context of cosmology, see \cite{NIEMEYER2020} for an overview. In very large volumes that fail to resolve $\ldb$, N-body simulations with FDM initial conditions adequately reproduce the suppression of small-scale clustering \cite{Irsic2017FirstSimulations,Armengaud2017,Schive2016CONTRASTINGDATA,Ni2019,Li2019}. Modified hydro solvers including a ``quantum pressure'' term motivated by the Madelung transformation of the Schrödinger equation have been used in the weakly nonlinear regime  \cite{Mocz2015,Veltmaat2016,Nori2018,Hopkins2018AMatter,Li2019}.However, only methods that solve the Schrödinger equation directly, capture the fully nonlinear wave-like dynamics in collapsed FDM structures, either in the entire computational domain \cite{Woo2009,Schive2014,Schwabe2016,Mocz2017} or in small subvolumes of a hybrid N-body-Schrödinger scheme \cite{Veltmaat2018}. 
Recently, the first hydrodynamical simulations with FDM including baryonic feedback on galactic scales were presented in Refs.~\cite{Mocz2019StarFilaments,Veltmaat2020}.

This paper introduces {\sc AxioNyx},\footnote{\url{https://github.com/axionyx}} a Schr\"{o}dinger-Poisson solver on adaptively refined regular meshes to facilitate detailed explorations of the astrophysical consequences of FDM. Built within the existing cosmology code {\sc Nyx} \cite{Almgren2013}, it inherits support for CDM and baryonic matter by means of a particle-mesh N-body scheme and a higher-order unsplit Godunov method for the gas dynamical equations, respectively. {\sc Nyx} itself is built upon AMReX, a powerful framework for block-structured adaptive mesh refinement (AMR) applications on massively parallel supercomputers \cite{Almgren2019}, boosting resolution in regions of interest and allowing the modeling of structure formation over many length scales. {\sc AxioNyx} implements both  pseudospectral  and  finite-difference methods for the Schr\"odinger-Poisson equation on the root grid, and uses a finite-difference method in refined regions. 

We examine the collapse of a spherical overdensity in a universe containing a mixture of cold and fuzzy dark matter as a test problem for {\sc AxioNyx}. This scenario is motivated by the many naturally light and weakly-interacting scalar particles predicted by string theory \cite{Arvanitaki2010}.  In our setup, they are represented by two dark matter components, one assumed to be sufficiently massive to be well-described by the standard N-body method and the other  exhibiting wave-like behavior governed by the Schrödinger-Poisson equation.

{\sc AxioNyx} is a cosmological solver but the Schr\"odinger-Poisson equation arises in several   contexts,  including boson stars \cite{Guzman2004,Schwabe2016,Mocz2017} and QCD axion miniclusters \cite{Eggemeier2019}.  {\sc AxioNyx} can easily be adapted to address these cases, and boson star condensation \cite{Levkov2018a} is one of the tests used to verify the code (see Appendix~\ref{appendix:iBScondensation}). Further, in many early-universe scenarios the post-inflationary universe can host the formation of transient, gravitationally collapsed overdensities, whose evolution is governed by the Schr\"odinger-Poisson equation  \cite{Musoke:2019ima,Niemeyer:2019gab}.

The paper is organized as follows. In Section~\ref{sec:method} we introduce our numerical methods, which are validated against a suite of test cases in the Appendices.   
We then investigate spherical collapse of MDM in Section~\ref{sec:collapse} and conclude in Section~\ref{sec:conclusions}. 

\section{Numerical Methods}
\label{sec:method}

The dynamics of FDM is well described by the Schr\"{o}dinger-Poisson equation
\begin{eqnarray}
    \label{eq:1part1}
     i\hbar\frac{\partial\psi}{\partial t} &=& -\frac{\hbar^{2}}{2ma^{2}}\nabla^{2}\psi+mV\psi\,,  \\
 \nabla^{2}V &=& \frac{4\pi G}{a}\rho \,  ,\quad\rho=|\psi|^{2} \, .  \label{eq:1part2}
\end{eqnarray}
Here, $\psi$ is the FDM wave function, $V$ denotes the gravitational potential, and $a$ is the scale factor. The constants $\hbar$, $m$, and $G$ stand for the Planck's constant, the mass of the axion, and Newton's gravitational constant, respectively.

We employ the {\sc Nyx}\footnote{https://github.com/AMReX-Astro/Nyx} code \cite{Almgren2013}, modified to handle FDM physics. {\sc Nyx} is an adaptive mesh refinement code with an MPI/OpenMP parallelization scheme, based on  the AMReX \cite{Almgren2019} library, that is known to scale well, even over thousands of distributed CPUs. It encompasses routines for hydrodynamics and an N-body solver for CDM physics. 

If periodic boundary conditions are applicable, which is often the case in cosmological scenarios, a pseudospectral solver is the fastest and most accurate algorithm to evolve the system of equations \ref{eq:1part1} and \ref{eq:1part2} on the Eulerian root grid. However, the refined regions will be aperiodic and we thus use a finite-difference scheme on subgrids.

The finite differencing scheme uses a fourth order Runge-Kutta solver to evolve equations~\ref{eq:1part1} and \ref{eq:1part2}, and had previously been integrated with \textsc{Nyx} \cite{Schwabe2016}. As an explicit algorithm, it suffers from stringent time-step constraints that scale quadratically with resolution, motivating the implementation of a pseudospectral solver.

\begin{table}
    \centering
    \begin{tabular}{c|cc|cc}
       $\alpha$ & \multicolumn{2}{c|}{2nd order} & \multicolumn{2}{c}{6th order} \\
        \hline
         &  $c_{\alpha}$& $d_{\alpha}$&  $c_{\alpha}$& $d_{\alpha}$\\
         1&0.5&1.0&0.39225680523878 &0.784513610477560\\
         2&0.5&0.0&0.51004341191846 &0.235573213359359\\
         3&&&-0.47105338540976 &-1.17767998417887\\
         4&&&0.06875316825251 &1.3151863206839023\\
         5&&&0.06875316825251 &-1.17767998417887\\
         6&&&-0.47105338540976 &0.235573213359359\\
         7&&&0.51004341191846 &0.784513610477560\\
         8&&&0.39225680523878&0.0
    \end{tabular}
    \caption{Weights used for higher order SP solver.}
    \label{tab:1}
\end{table}

\begin{figure*}[tb]
    \centering
    \includegraphics[width=\linewidth]{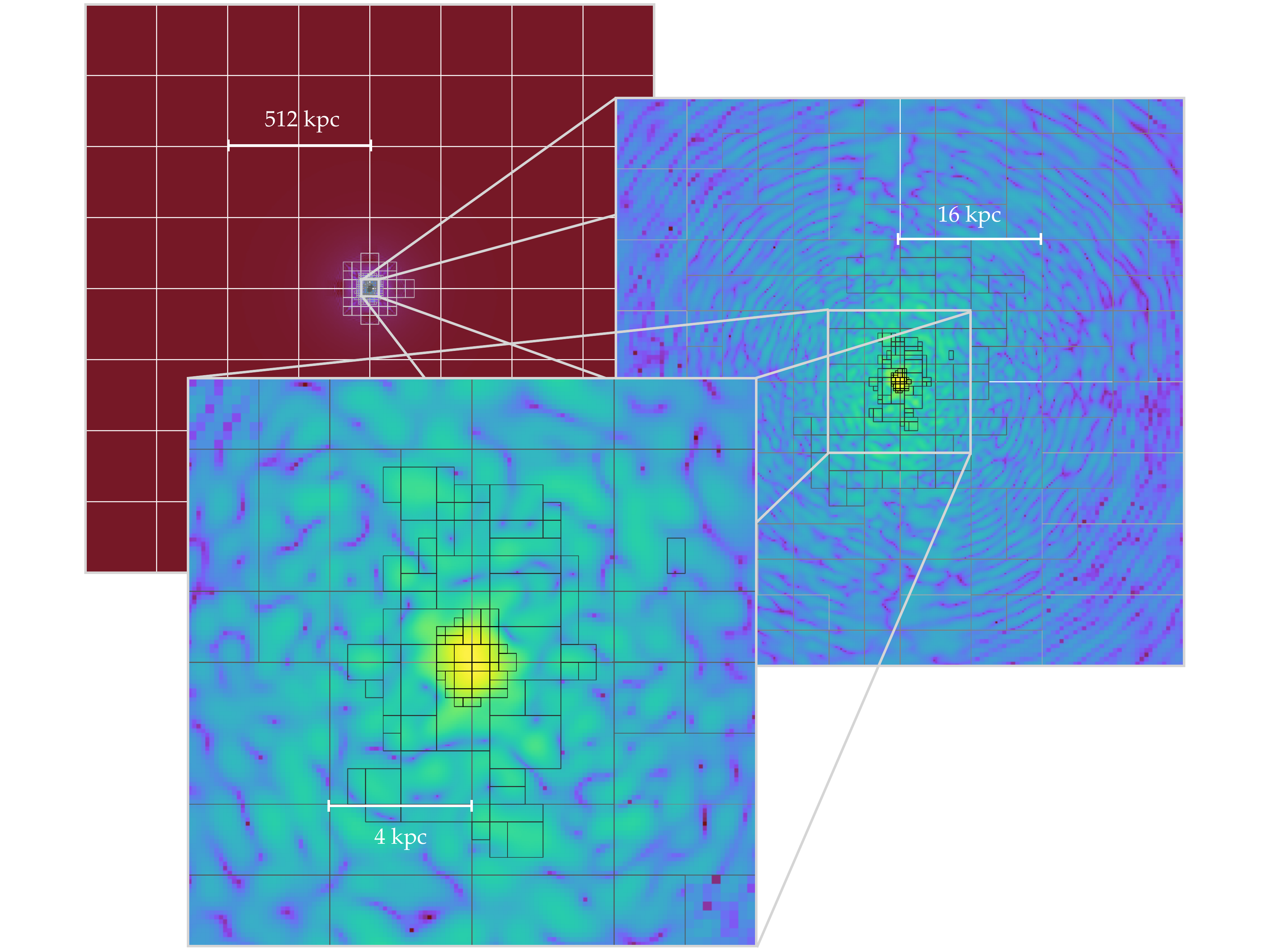}
    \caption{Slice through the final FDM density field  with 30\% FDM and 70\% CDM. The central region is highly refined, and resolves the FDM granules which form after shell crossing. Rectangular boxes show regions with the same level of refinement -- the box edges are coloured such that they are darker at higher levels of refinement.}
    \label{fig:zoomin}
\end{figure*} 

The public  AMReX\footnote{https://github.com/AMReX-Codes/amrex} repository already includes the stand-alone version of HACC's  distributed-memory, pencil-decomposed, parallel 3D FFT routines and wrappers linking those routines to the  AMReX framework.\footnote{https://xgitlab.cels.anl.gov/hacc/SWFFT} We optimized the HACC libraries for OpenMP parallelization and obtain good weak scaling at up to $512$ cores, quantified by the negligible dependence of the solution time with respect to the number of processors for a fixed problem size per processor. We implement two different pseudospectral schemes, one at second order and the other at sixth order. The second order scheme matches that implemented in {\sc PyUltraLight}\footnote{https://github.com/auckland-cosmo/PyUltraLight}~\cite{Edwards:2018ccc}. The sixth order scheme replicates the method described in
Ref~\cite{Levkov2018a}

\begin{align}
    \label{eq:1a}
    \psi(t+\Delta t) = \prod^{\alpha}e^{-imd_{\alpha}\Delta tV_{\alpha}(x)/\hbar}e^{-i\hbar c_{\alpha}\Delta t k^{2}/2ma^{2}}\psi(t)\, ,
\end{align}
with weights summarized in \autoref{tab:1}. Other higher-order algorithms could easily be implemented.

Eulerian grids used to deposit information from N-body particles in cloud-in-cell simulations appropriate for CDM are typically only adaptively refined in over-dense regions to correctly capture the strong dynamics in filaments and halos. Conversely, in FDM simulations  the velocity is inferred from the gradient of the wave function's complex phase; even low density regions may have to be refined if they contain high-speed flows.  Consequently,  we employ the L\"ohner error estimator \cite{Loehner1987}, which is also implemented in Enzo \cite{Bryan2014} and {\sc FLASH} \cite{Fryxell2000} and was also used with FDM cosmological simulations in Refs~\cite{Schive2014,mina2019scalar}. However, as in Ref.~\cite{Schive2014}, we prohibit grid refinement in regions below a certain density, chosen so that their detailed behaviour does not significantly alter the overall dynamics. 

The power of adaptive refinement is demonstrated in \autoref{fig:zoomin}, which shows the final snapshot of a spherical collapse with 30\% FDM and 70\% CDM. Six levels of refinement were used, increasing the resolution by a factor of 2 with each level. 

\begin{figure}
    \centering
    \includegraphics[width=\linewidth]{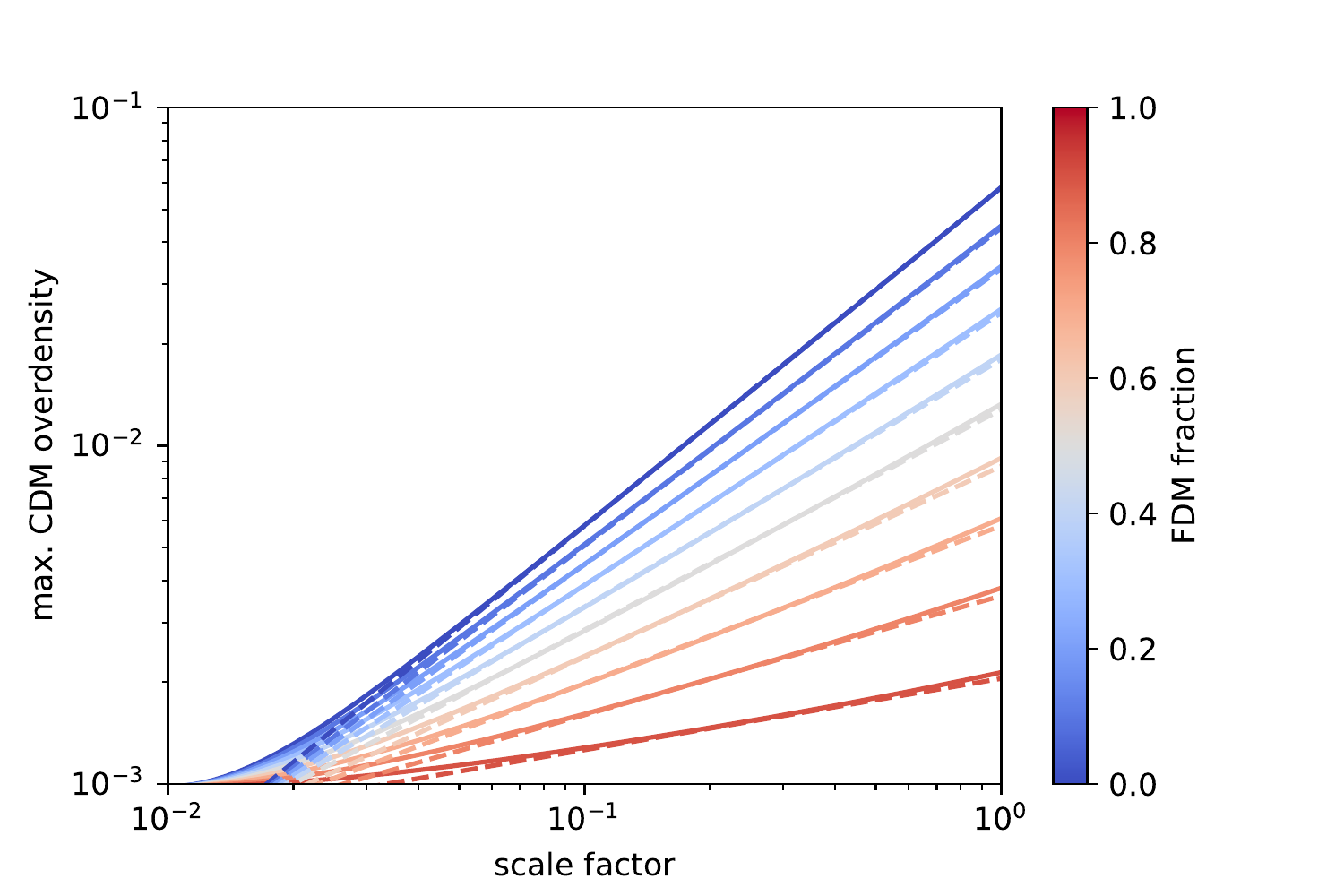}
    \caption{Evolution of the maximum CDM over-density during spherical collapse for different CDM fractions $(1-f)$. The numerically obtained power law growth (solid lines) is well approximated by \autoref{eq:cdmgrowth} (dashed lines).}
    \label{fig:8}
\end{figure}
 
We conducted a number of tests of the validity and accuracy of {\sc AxioNyx}, as described in the Appendix. These begin with comparisons of the pseudospectral Schr\"odinger solver with static potentials for which exact solutions are known, verifying the scaling with timestep size. In \ref{appendix:isolatedSoliton} we verify that isolated solitons in a static background behave as expected, showing small departures from perfectly static solutions on small grids. In \ref{appendix:iBScondensation} we replicate simulations of boson star condensation from random Gaussian initial conditions \cite{Levkov2018a}.  Since all regions are equally important, this problem is particularly suitable for a pseudospectral solver which allows the use of significantly larger time steps and lower spatial resolution than finite difference methods.

The cosmological dynamics are verified by following the linear evolution of a one-dimensional, self-gravitating, sinusoidal over-density as it crosses the FDM Jeans scale in a comoving box (\ref{sec:lme}). Finally, in \ref{sec:sphcol} we investigate  support against spherical gravitational collapse near the Jeans scale. This setup is useful to test the L\"ohner refinement criterion and subcycling; even extreme cases with more than 100 sub-steps between root and first level yields well converged results.  Results from a representative simulations are shown in Figure~\ref{fig:zoomin}.

\begin{figure}
    \centering
    \includegraphics[width=\linewidth]{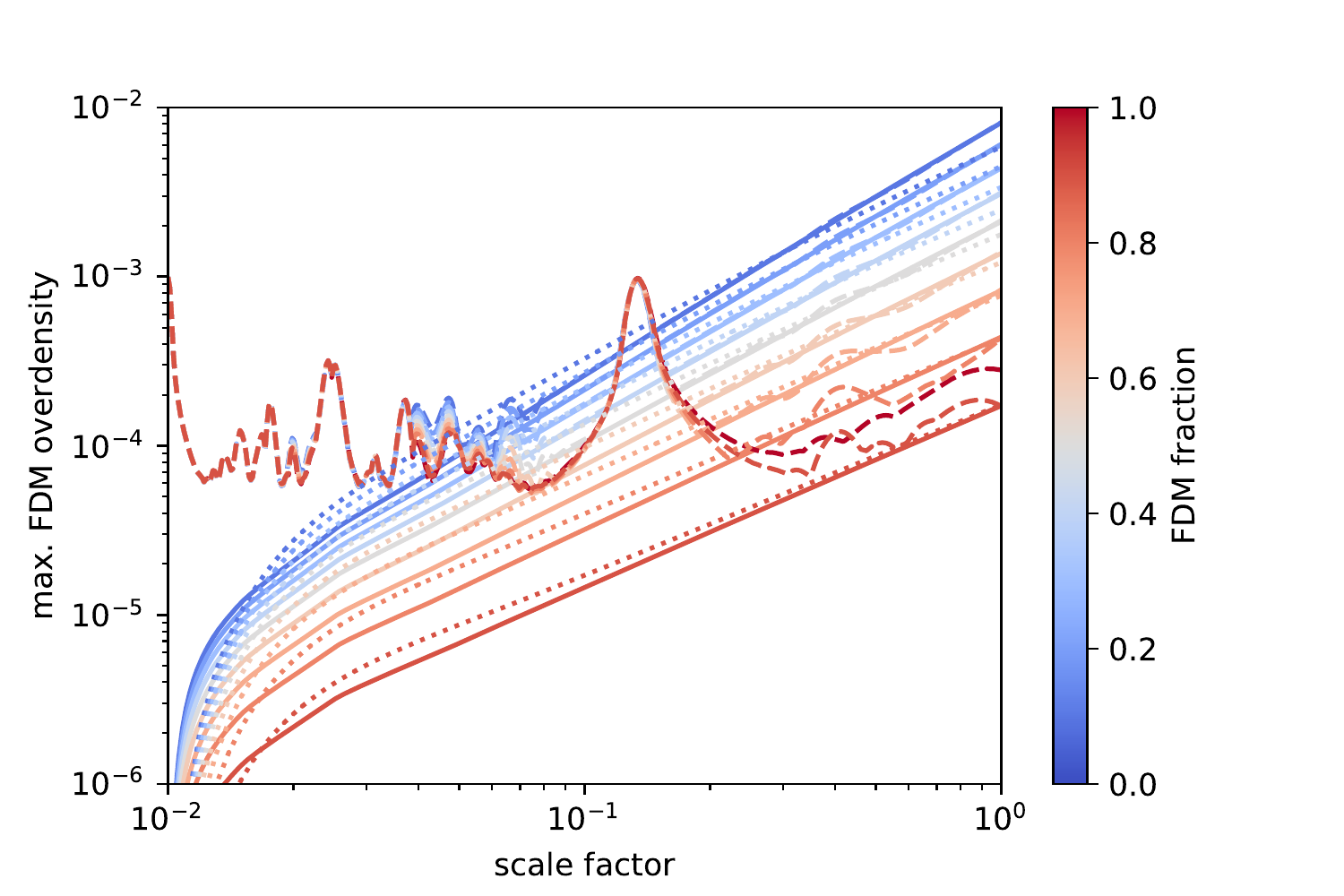}
    \caption{Evolution of the maximum FDM over-density during spherical collapse for different $f$. The initially homogeneous FDM fields (solid line) contract according to linear theory (dotted lines). The power law growth at low redshifts is independent of FDM initial conditions as small initial FDM over-densities (dashed line) show the same late time behaviour.}
    \label{fig:7}
\end{figure}

\section {Spherical collapse of mixed dark matter}
\label{sec:collapse}

In this section we extend the spherical collapse to a MDM scenario comprised of FDM and CDM with the fraction of FDM defined as
\begin{equation}
f=\frac{\rho_{{\rm FDM}}}{\rho_{{\rm FDM}}+\rho_{{\rm CDM}}}.
\end{equation}
For FDM, we use the sixth order Schr\"{o}dinger-Poisson solver on the root grid and the finite difference algorithm on higher levels, for the CDM we make use of the N-body scheme as implemented in \textsc{Nyx}. 

\subsection{The linear regime}

The linear evolution of MDM is governed by the following system of coupled differential equations:
\begin{widetext}
\begin{subequations}
\begin{align}
    \ddot{\delta}_{\rm FDM}&+2H\dot{\delta}_{\rm FDM}+\left(\frac{k^{4}\hbar^{2}}{4m^{2}a^{4}}-4\pi Gf\overline{\rho} \right)\delta_{\rm FDM} = 4\pi G(1-f)\overline{\rho}\delta_{\rm CDM}\, ,\label{eq:lingr2a}\\
    \ddot{\delta}_{\rm CDM}&+2H\dot{\delta}_{\rm CDM}-4\pi G(1-f)\overline{\rho}\delta_{\rm CDM} = 4\pi Gf\overline{\rho}\delta_{\rm FDM}\, .\label{eq:lingr2b}
\end{align}
\end{subequations}
\end{widetext}

\begin{figure*}
        \centering
    \includegraphics[width=\linewidth]{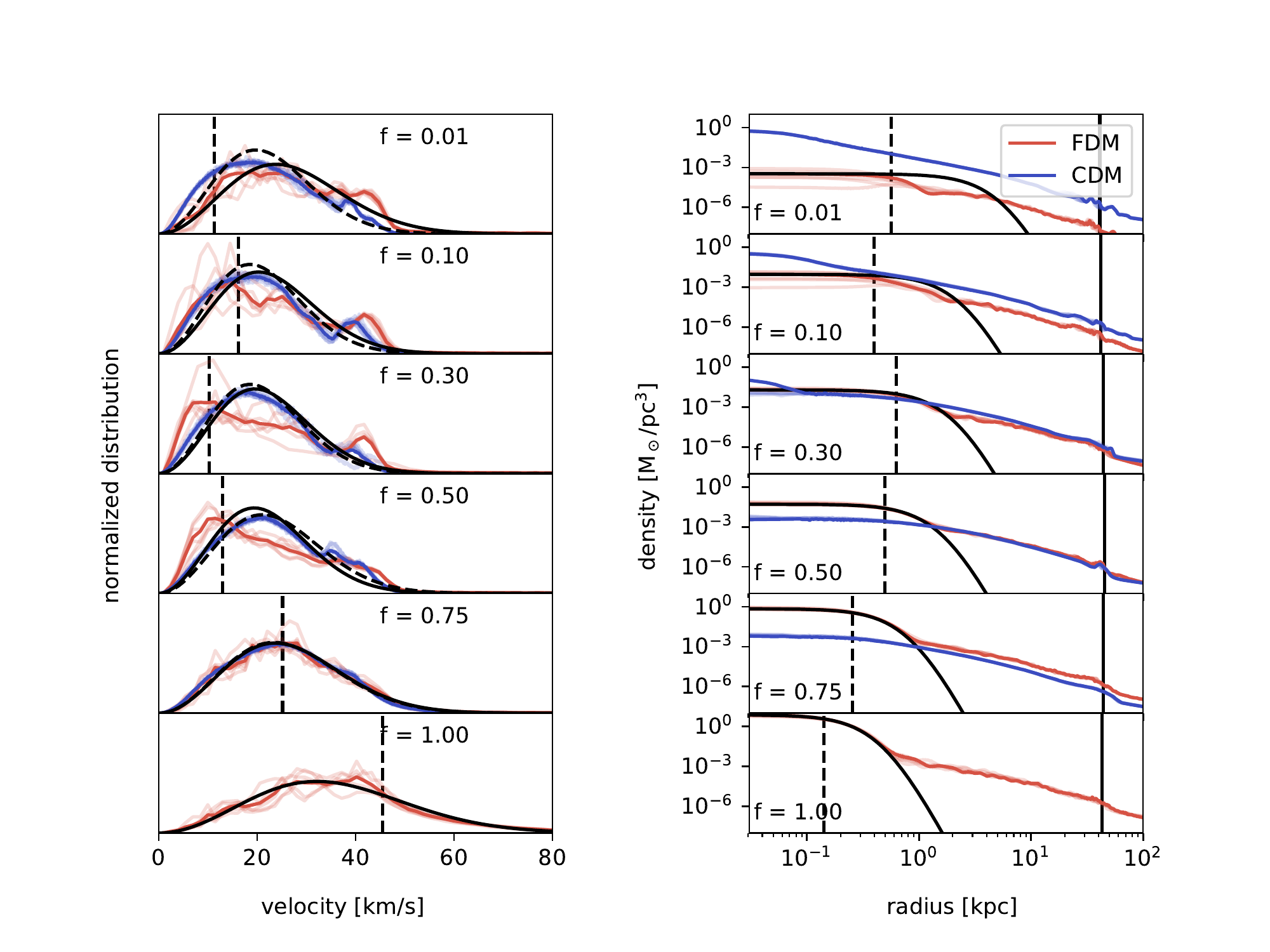}
    \caption{(Left) Normalized velocity spectra of the FDM and CDM components for various $f$. For large $f$ they are well fitted by Maxwell-Boltzmann distributions shown in black. Dashed horizontal lines indicate soliton velocities as defined in \autoref{vc}. (Right) Radial density profiles of the FDM and CDM components for various $f$. For large $f$ central FDM profiles are well fit by solitonic profiles shown in black. Dashed vertical lines indicate soliton radii while solid vertical lines represent virial radii.}
    \label{fig:velprof}
\end{figure*}

We choose a box size an order of magnitude below the FDM Jeans scale. Accordingly, the growth of any local deviation from the critical density is highly suppressed in the FDM sector. It is therefore reasonable to start with $\delta_{\rm FDM}(z_{\rm ini})=0$. In contrast, we introduce a Gaussian overdensity with maximum amplitude $\delta_{\rm CDM}(z_{\rm ini})=0.001$ on top of a CDM background at an initial redshift $z_{\rm ini}=99$. The time evolution of $\delta_{\rm CDM}$ is shown in \autoref{fig:8}. As expected, it is suppressed relative to a pure CDM simulation \cite{Hu1998}, via
\begin{align}
\label{eq:cdmgrowth}
    \delta_{\rm CDM}(a) \propto a^{(\sqrt{1+24(1-f)}-1)/4}\, .
\end{align}

\autoref{eq:cdmgrowth} can be obtained by setting $\delta_{\rm FDM}=0$ in \autoref{eq:lingr2b}. In \autoref{fig:7} we show the numerically obtained maximum FDM overdensity. It is well approximated by the solution of \autoref{eq:lingr2a} for $\delta_{\rm CDM}(a)$ as given in \autoref{eq:cdmgrowth}. The late time evolution of modes $k\gg k_{J}(a)$ follows a single power law
\begin{align}
\label{eq:fdmgrowth}
    \delta_{\rm FDM}(a) \propto a^{(\sqrt{1+24(1-f)}+3)/4}\, .
\end{align}

We verify that the growth is not an artifact of the homogeneous FDM initial conditions by running the same simulation with an added FDM overdensity with initial maximum amplitude $\delta_{\rm FDM}(z_{\rm ini})=\delta_{\rm CDM}(z_{\rm ini})=0.001$. As seen in \autoref{fig:7}, the initial overdensity rapidly disperses. The emerging fluctuations only start to grow once the potential well formed by the increasing CDM overdensity is deep enough and exhibit the same polynomial growth as found previously.

\subsection{The non-linear regime}

Employing adaptive mesh refinement with a $1024{}^{3}$ root grid and up to $6$ levels of refinement, we investigate spherical collapse all the way into the highly non-linear regime after shell crossing. We ran simulations with $11$ different mixed dark matter fractions $f$ in a $2\,$Mpc comoving box. The typical mass of the collapsed halo is $3.5\times 10^9$ solar masses. The AMR structure of one of our simulations can be seen in \autoref{fig:zoomin}, which depicts slices through the final FDM density. Focusing on the central halo region we see the expected granular structure in FDM as a result of interfering plane waves. 

In \autoref{fig:velprof} we show velocity distributions on the left side and density profiles on the right. Due to FDM exhibiting oscillations in time, we show six different snapshots, extracted from the final stages of our simulations. They are shown in transparent blue or red lines for CDM and FDM, respectively. The thick red (blue) lines show profiles averaged over time.

\begin{figure}[tb]
    \centering
    \includegraphics[width=\linewidth]{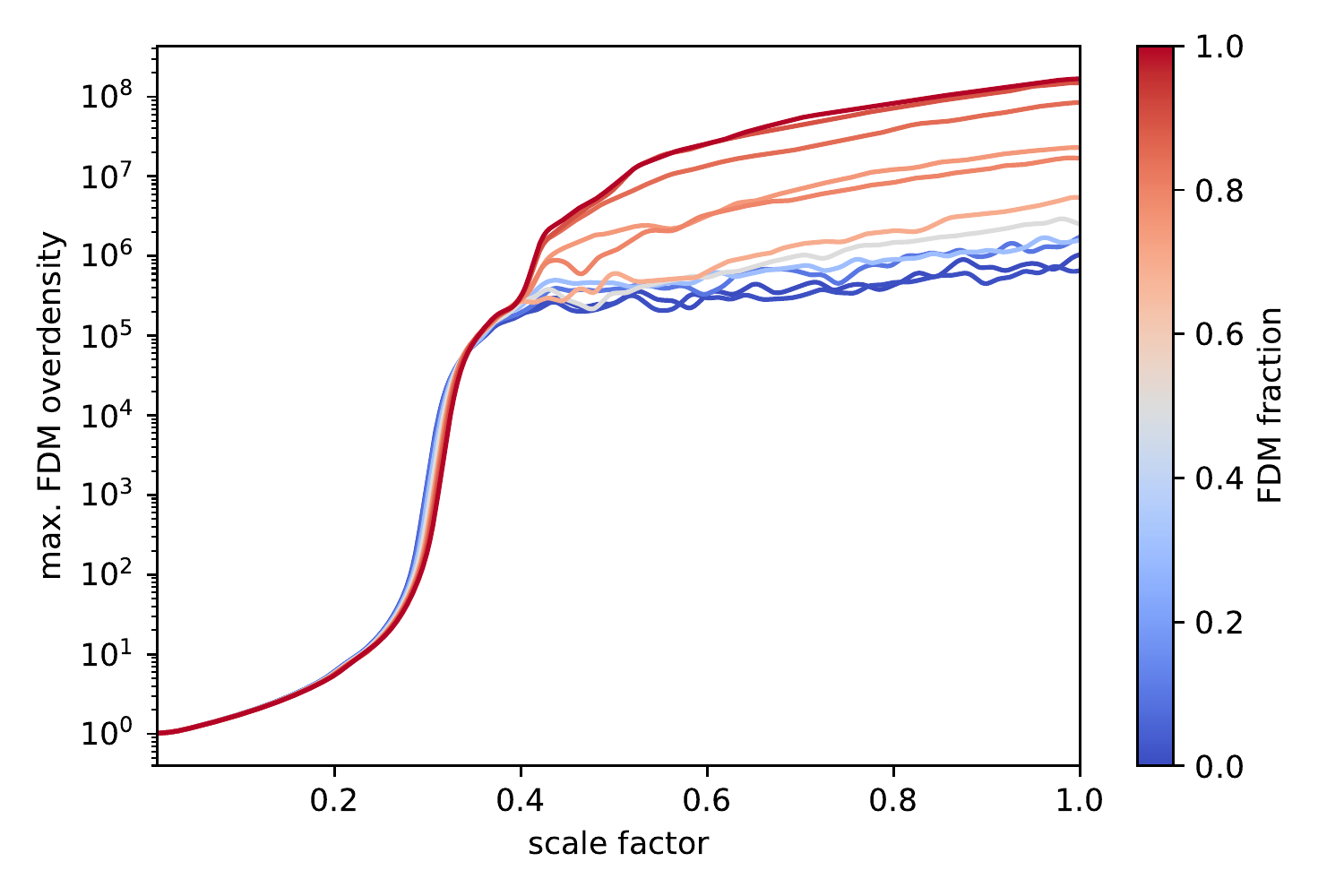}
    \caption{Maximum central FDM densities normalized by initial values. Soliton formation breaks the degenerate evolution around a scale factor $a=0.4$ corresponding to roughly one and a half free-fall time.}
    \label{fig:maxdens}
\end{figure}

In the outer NFW-like halo $f$ remains constant even after non-linear collapse and velocity spectra are only mildly sensitive to different $f$. This demonstrates that the Schr\"odinger-Vlasov correspondence also holds for mixed dark matter, i.e. that FDM averaged over multiple de Broglie wavelengths behaves as CDM. This excludes formation of solitons which have no CDM analog. The intriguing conclusion is that analytic estimates of condensation \cite{Levkov2018a}, heating and cooling \cite{Niemeyer2019}, and relaxation processes \cite{Hui2017} in the FDM sector which depend on the effective granule mass can be straightforwardly generalized to mixed dark matter scenarios by re-scaling the effective granule mass by $f$.  

Within the central region of a pure FDM halo, simulations reveal the formation of solitonic ground state solutions. These solutions are present only if $f \gtrsim 0.1$. In this case the inner part of the FDM radial density profiles can be well approximated by a soliton profile as depicted in \autoref{fig:velprof}. For lower $f$ we see strong fluctuations in the central FDM density profiles, which cannot be fitted by the 
original
or the modified soliton profiles \cite{Veltmaat2020}. We thus conclude that  for mixed dark matter soliton formation only occurs if $f \gtrsim 0.1$. In that case the central density $\rho_c$ significantly increases beyond the initial collapse, breaking its initially degenerate evolution, as seen in \autoref{fig:maxdens}. 

\begin{figure}
    \centering
    \includegraphics[width=\linewidth]{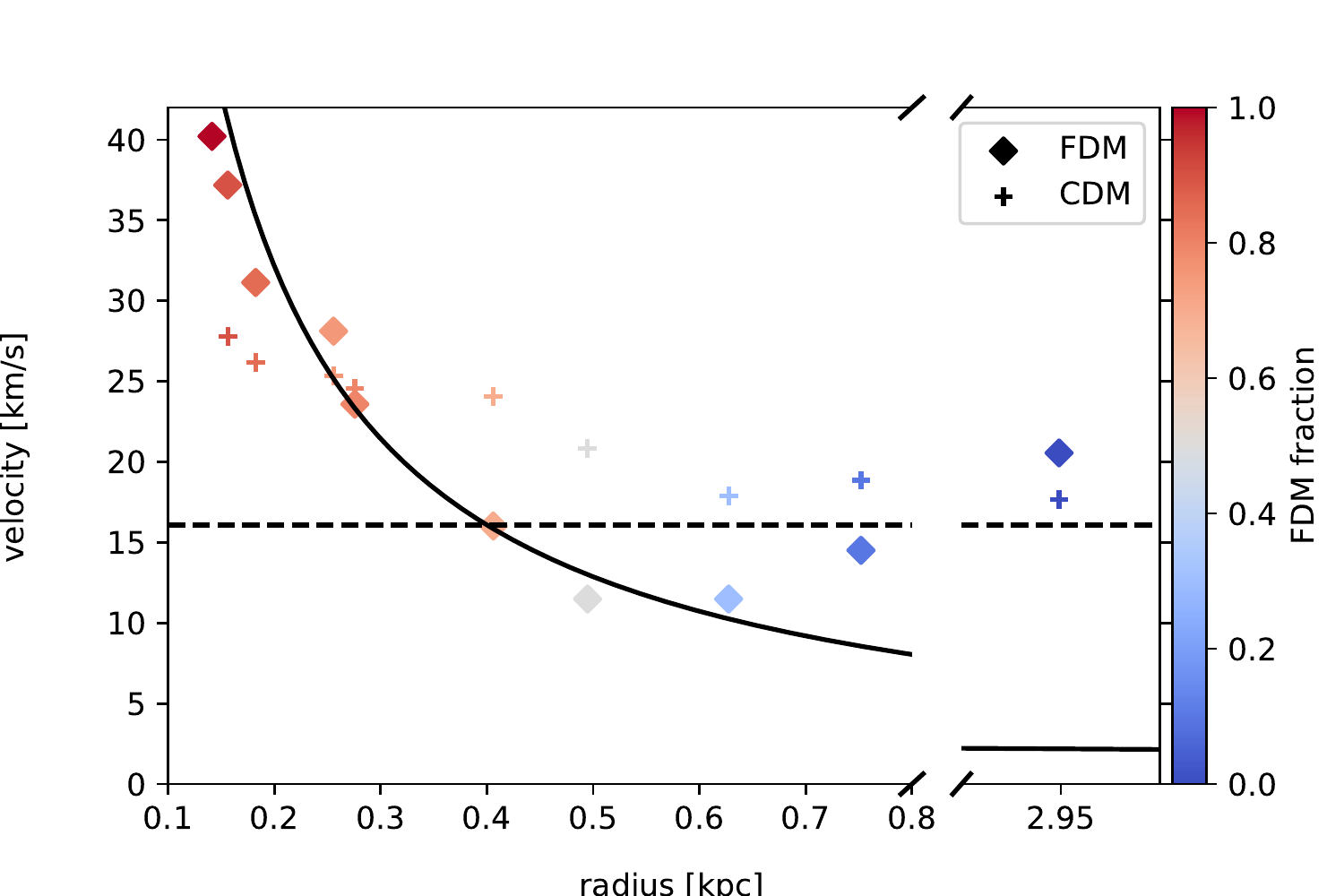}
    \caption{Peaks of the final FDM (diamonds) and CDM (crosses) velocity spectra in the inner halo region as presented in \autoref{fig:velprof} as a function of soliton radii for different $f$. FDM velocitiy peaks tightly follow \autoref{vc} indicated by the solid black line. For comparison, the dashed line indicates the halo's virial velocity.}
    \label{fig:vel}
\end{figure}

The vertical dashed lines on  the right side of \autoref{fig:velprof} indicate the soliton radius $r_c$ at which the spherically averaged FDM density drops to half its central value. The corresponding soliton velocities \cite{Mocz2017}
\begin{align}
    \label{vc}
    v_c=\frac{2\pi}{7.5}\frac{\hbar}{mr_c}
\end{align}
are represented by the dashed lines on the left. They align well with the peaks of the Maxwell-Boltzmann-like distributed, normalized FDM velocity spectra \cite{Veltmaat2018}
\begin{align}
    f(\textbf{v}) = \frac{1}{N}\left|\int\text{d}^3 x\exp\left[-im\textbf{v}\cdot\textbf{x}/\hbar\right]\psi(\text{x})\right|^2\, .
\end{align}
We integrated the final FDM state in the central $8$ kpc cubed box, which encompasses at least dozens of granules apart from the soliton. The tight correlation between maxima in FDM velocity spectra and soliton velocities shown in \autoref{fig:vel} suggests that the soliton is in kinetic equilibrium with its surrounding. During soliton formation central velocities both in the FDM and CDM component decouple from the virial velocity of the halo indicated by the dashed, horizontal line in \autoref{fig:vel}. In contrast, in simulations with insufficient FDM content for soliton formation ($f\lesssim 0.1$), both the CDM and FDM velocity spectra peak close to the  virial velocity of the halo. 

\begin{figure}[tb]
    \centering
    \includegraphics[width=\linewidth]{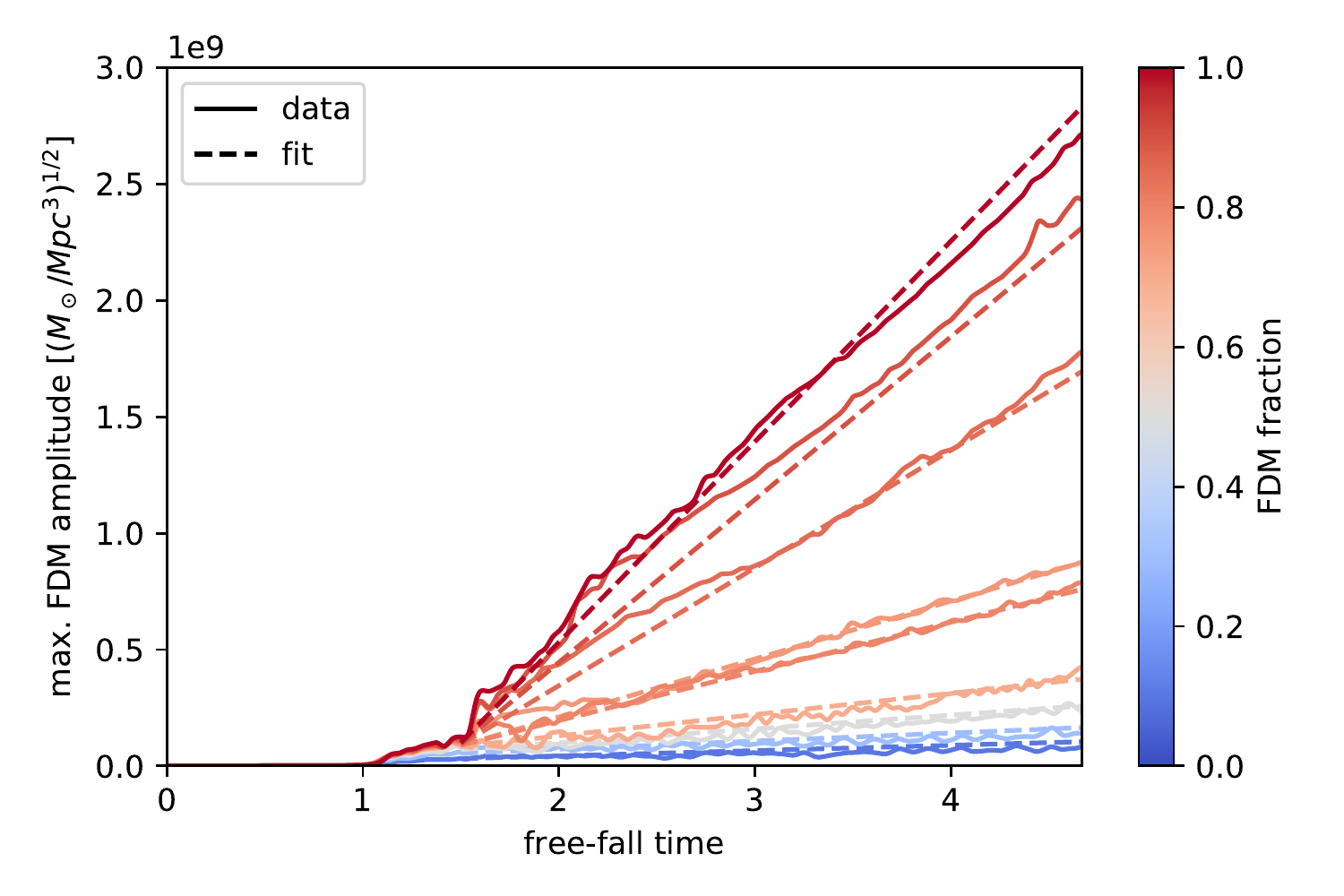}
    \caption{Linear growth of maximum FDM amplitudes well described by \autoref{lingr}.  }
    \label{fig:maxamp}
\end{figure}

As can be seen in \autoref{fig:maxamp}, the maximum soliton amplitude
\begin{align}
    \label{lingr}
    A(t) = A_{1}\cdot (t-t_{0})/\tau_{\text{gr}}+A_{0}f^{1/2}
\end{align}
with
\begin{align}
    \label{taugr}
    \tau_{\text{gr}} = \frac{0.7\sqrt{2}}{12\pi^3}\frac{m^3v_c^6}{G^2\rho_{c}^2\Lambda}\simeq 0.015 \frac{t_c}{\Lambda}
\end{align}
grows linearly in time. Here, $\Lambda=\log(r_{\text{vir}}/r_c)$ is the Coulomb logarithm, $r_{\text{vir}}$ the virial radius, and in the last step we used soliton identities as listed in the Appendix B of \cite{Hui2017}. Numerically, we found $A_{1}=1350\, (M_\odot/\text{Mpc}^3)^{1/2}$, $A_{0}=10^8\, (M_\odot/\text{Mpc}^3)^{1/2}$, and $t_{0}=1.5t_{\text{ff}}$ with free-fall time $t_{\text{ff}}=0.003$\,s Mpc/km. Since $v_c$ only mildly depends on the FDM fraction while density scales linearly with $f$, growth in soliton amplitude is suppressed roughly quadratically with $f$.  

\section{Conclusions}
\label{sec:conclusions} 

We developed the highly-parallelised numerical code {\sc AxioNyx} for simulations of self-gravitating mixed fuzzy and cold dark matter. The code features adaptive mesh refinement which allows tremendous effective resolution in regions of interest, while remaining relatively cheap on the largest scales. We used a 6th order pseudospectral solver on the root grid and a 4th order finite-difference solver on the refined subgrids. We employ the Poisson solver available in {\sc Nyx}. An exhaustive test suite of progressively more complex problems demonstrates the correctness and efficiency of {\sc AxioNyx} and highlights its vast range of potential applications.

With mixed dark matter spherical collapse simulations we show that below the Jeans scale fuzzy dark matter overdensities $\delta_{\rm FDM}(a)$ are supported against gravitational collapse by gradient energy. In turn, cold dark matter collapse is increasingly suppressed with higher fractions of fuzzy dark matter. The deepening gravitational potential shrinks the Jeans scale and results in late time fuzzy dark matter collapse. We provide analytic estimates for the evolution of $\delta_{\rm FDM}(a)$ and $\delta_{\rm CDM}(a)$ in the linear regime.

Adaptively refined spherical collapse simulations well in the non-linear regime after shell-crossing confirm that the Schr\"odinger-Vlasov correspondence holds even in the case of mixed dark matter. Averaging over multiple de Broglie wavelengths we see a confluent
evolution of cold and fuzzy dark matter due to both species responding to the same gravitational potential.
Analytic estimates of fuzzy dark matter soliton condensation, heating and cooling, and relaxation processes all depend on the effective granule mass. They can thus be straightforwardly generalized to mixed dark matter scenarios by re-scaling the effective granule mass by the fuzzy dark matter fraction.

The degeneracy between FDM and CDM is broken by 
the soliton formation which occurs in the center of collapsing overdensities if the fraction of fuzzy dark matter is sufficiently large, $f\gtrsim 0.1$. While staying in kinetic equilibrium, the linearly growing soliton heats up its surrounding inner halo region.

Our adaptively refined simulations demonstrate that cosmological simulations with an effective resolution $\sim 10^5$ are feasible with \textsc{AxioNyx}. Assuming maximum velocities of $\sim 100\,$km/s comparable to those in \autoref{fig:velprof} and a FDM mass $\sim 10^{-22}\,$eV, we are able to resolve the minimum de Broglie wavelength with at least $6$ cells when allowing for grid resolutions $\sim 100\,$pc in a $\sim 10\,$Mpc comoving box.

\acknowledgements

We thank Benedikt Eggemeier, Oliver Hahn, Shaun Hotchkiss, Emily Kendall, Doddy Marsh, Nathan Musoke, and Jan Veltmaat for important discussions, and the AMReX code development team, especially Ann Almgren and Guy Moore, for their assistance.    Computations described
in this work were mainly performed with resources provided by the North-German Supercomputing Alliance (HLRN). Additionally, the authors wish to acknowledge the use of New Zealand eScience Infrastructure
(NeSI) high performance computing facilities, consulting support and/or training services as part of this research. New Zealand’s national facilities are provided by NeSI and funded jointly by NeSI’s collaborator institutions and through the Ministry of Business, Innovation \& Employment's Research Infrastructure programme. We acknowledge the \textsc{yt}~\cite{2011ApJS..192....9T} toolkit that was used for the analysis of numerical data. 
BS acknowledges support by the Deutsche Forschungsgemeinschaft. JCN acknowledges funding by a Julius von Haast Fellowship Award provided by the New Zealand Ministry of Business, Innovation and Employment and administered by the Royal Society of New Zealand. We acknowledge support from the Marsden Fund of the Royal Society of New Zealand.

\bibliography{axionyx}

\appendix

\begin{figure}[tb]
    \centering
    \includegraphics[width=\linewidth]{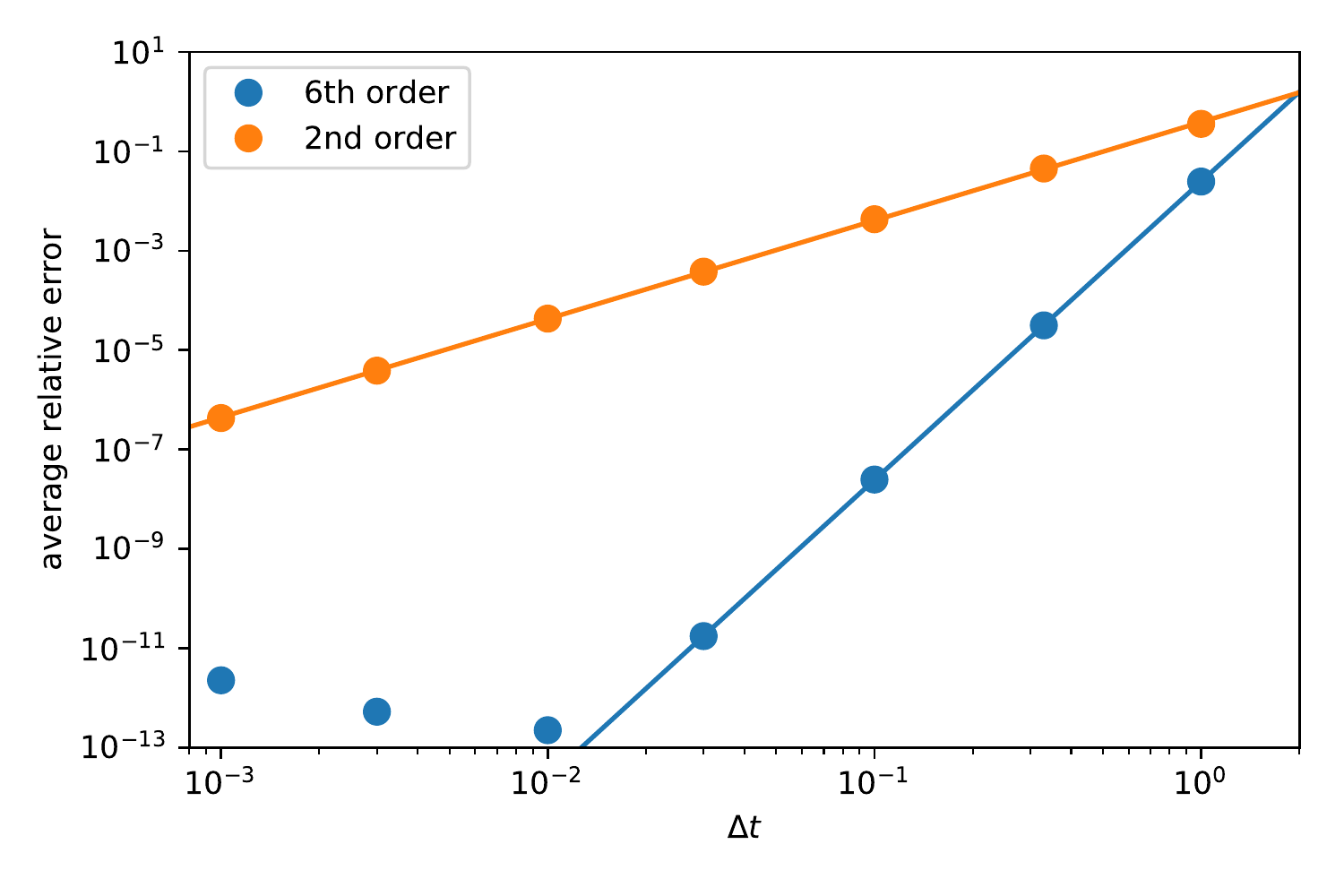}
    \caption{Spatially averaged error of the final state as a function of the time step for the two solvers. Each point is a separate simulation with a fixed time step. The fitted slopes of the line accurated reproduce the expected scalings. At sixth order with $\Delta t > 10^{-2}$ the error is dominated by numerical noise due to finite precision floating point numbers.}
    \label{fig:2vs6}
\end{figure}

\begin{figure}[tb]
    \centering
    \includegraphics[width=\linewidth]{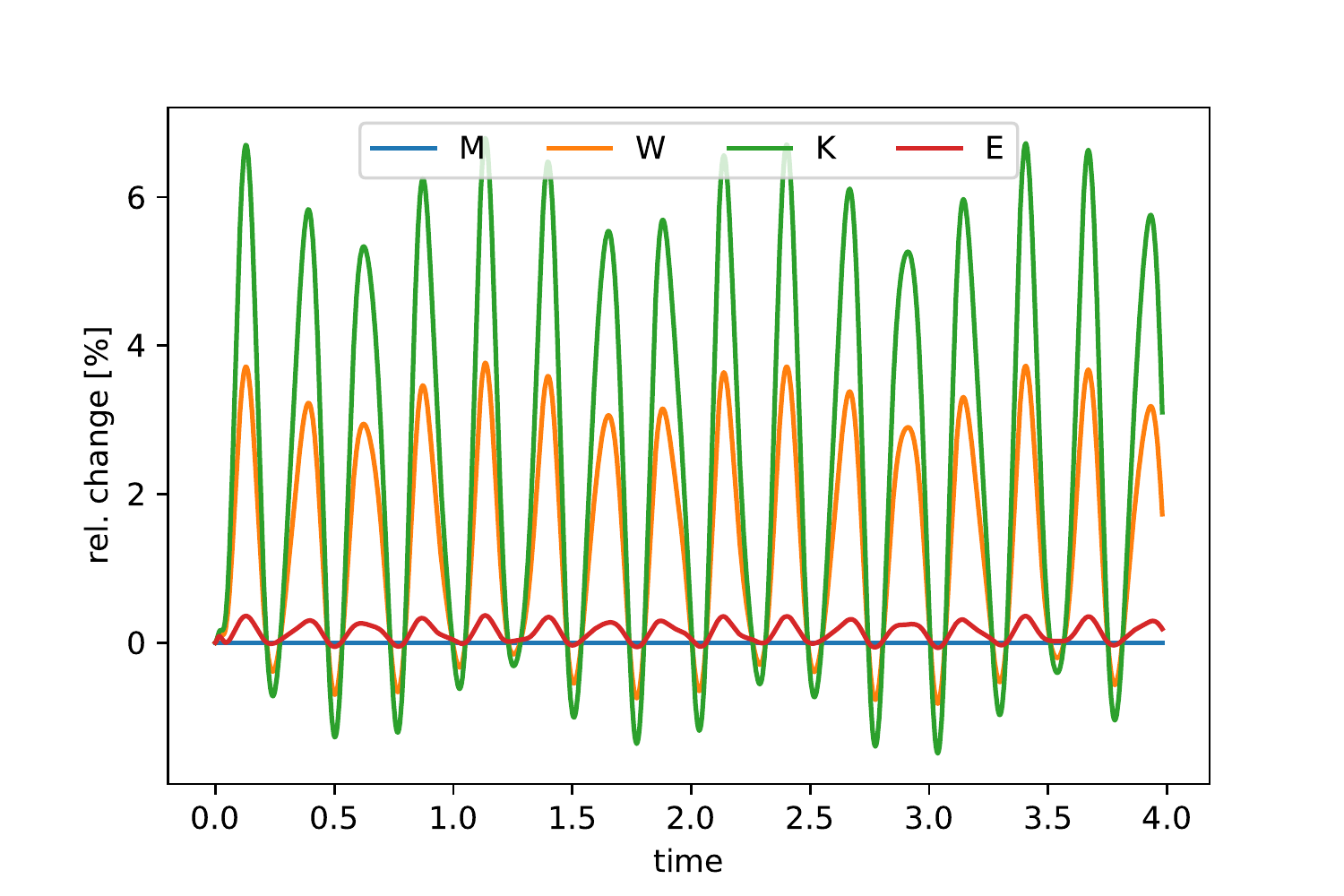}
    \caption{Relative changes of global quantities over time during the simulation of a single soliton. Due to discreteness errors, it is slightly ringing with a mass dependent frequency as can be inferred from the potential energy $W$ and kinetic energy $K$ (see, e.g., \cite{Schwabe2016} for the definition of conserved quantities). Total mass and energy are well conserved even in the depicted low resolution run.}  
    \label{fig:3}
\end{figure}

\begin{figure}[b]
    \centering
    \includegraphics[width=\linewidth]{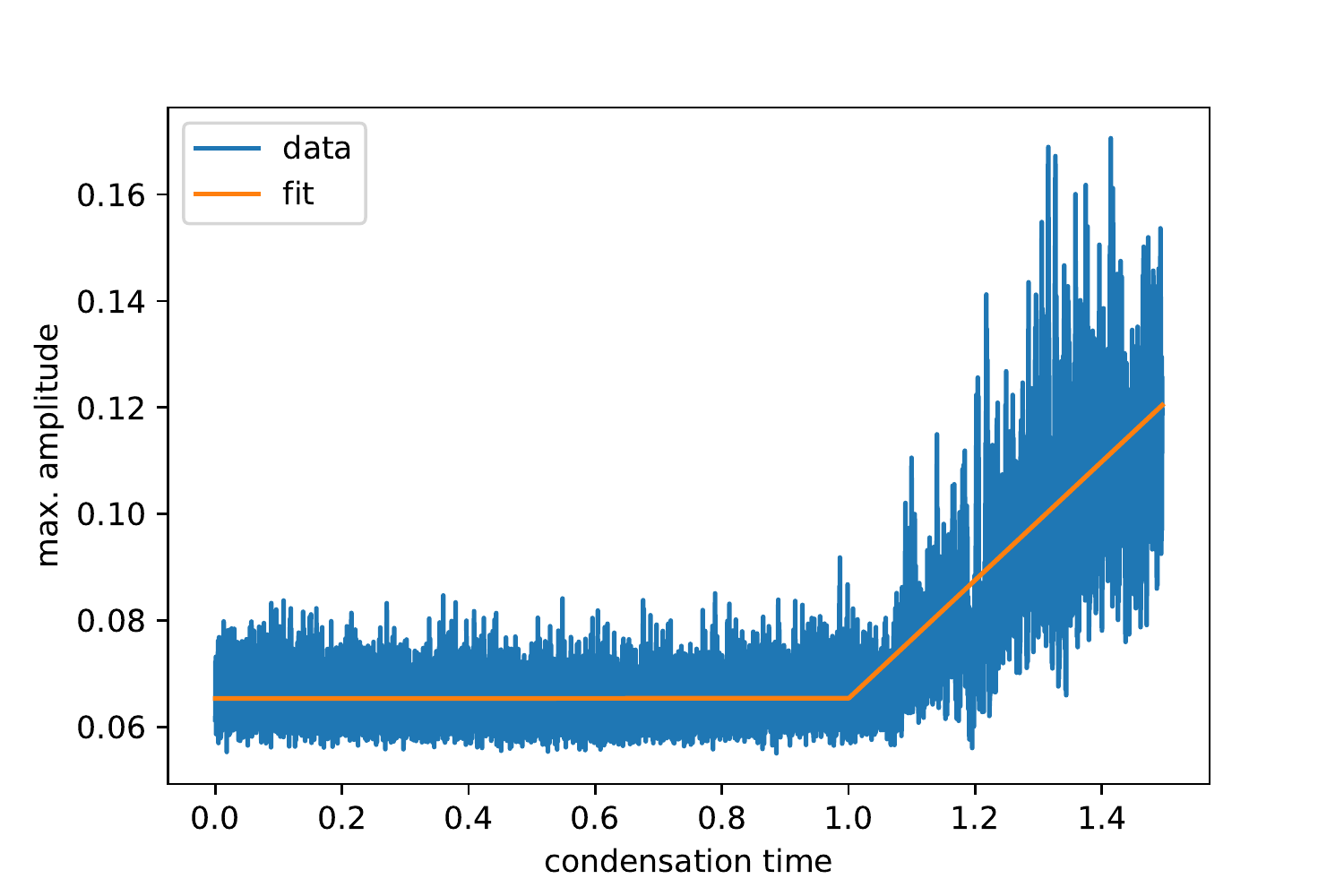}
    \caption{Linear growth of the maximum amplitude due to the condensation of a Boson star. \label{fig:2}}

     \includegraphics[width=\linewidth]{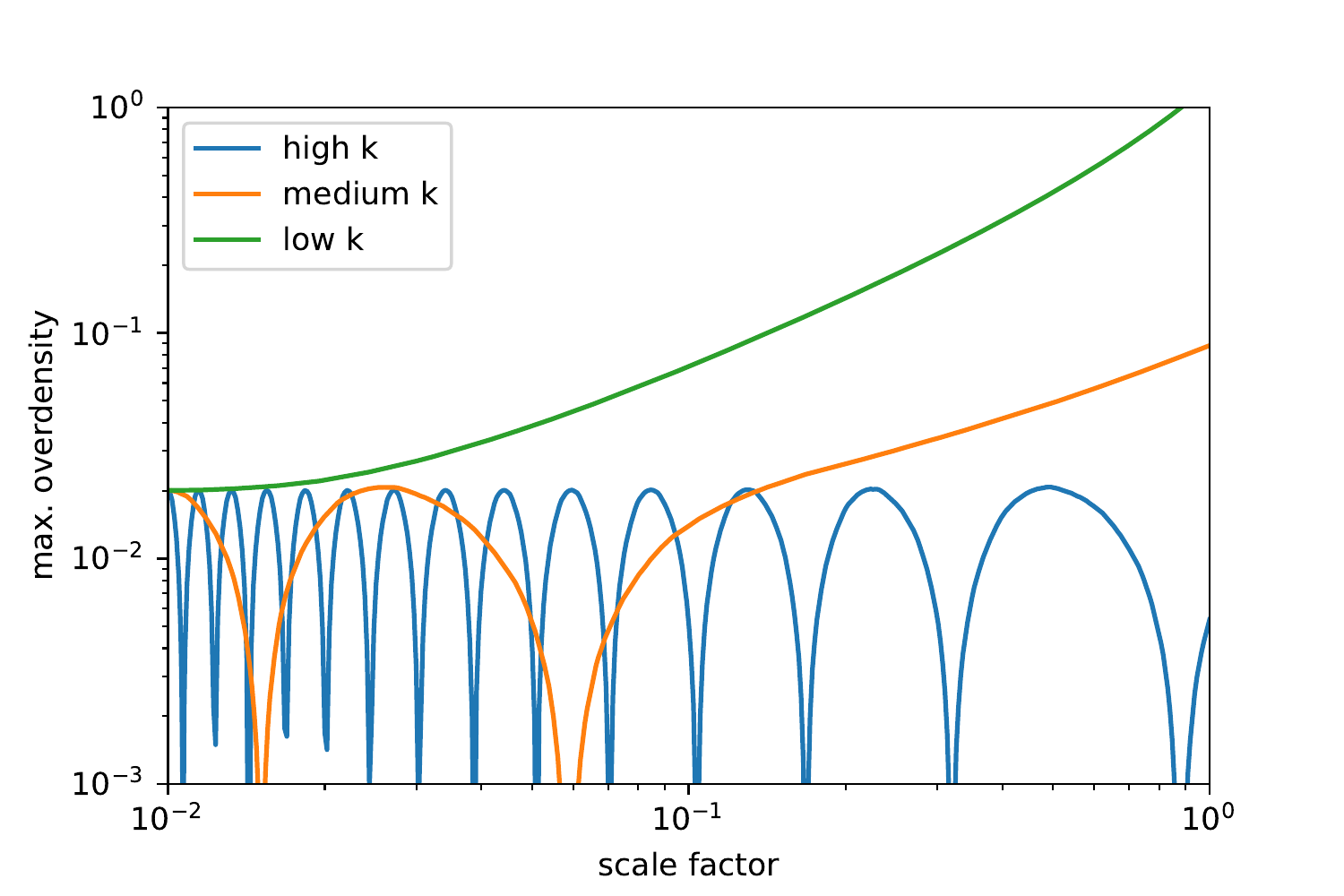}
    \caption{Time evolution of different self-gravitating modes in an expanding background. Modes with wave numbers larger than the Jeans scale are stabilized by wave coherence effects. \label{fig:5}}

\end{figure}

\section{Code Verification}

We  tested the code with a suite of increasingly challenging computations. 
\subsection{Static potential tests}
\label{appendix:staticPotential}
Setting the gravitational potential $V=0$, an initially Gaussian FDM overdensity broadens via diffusion.  The dynamics can be solved exactly,
\begin{align}
    \rho(x,t) &= \frac{1}{\sqrt{2\pi\sigma^{2}(t)}}\exp\left[-\frac{(x-\overline{x}(t))^{2}}{2\sigma^{2}(t)}\right]\, ,\\
    \sigma^{2}(t) &= \sigma_{0}^{2}\left[1+\left(\frac{(t-t_{0})\hbar}{2m\sigma_{0}^{2}}\right)^{2}\right]\, ,
\end{align}
where the initial width $\sigma(t_{0})=\sigma_{0}$ and $\overline{x}(t)=0$. The maximum relative errors are smaller then $10^{-12}$ until $\sigma(t_{f})=2\sigma_{0}$ even for a low resolution grid spacing, $\Delta x = 0.5\sigma_{0}$. Placing the Gaussian over-density in an external harmonic oscillator potential
\begin{align}
    V(x) = -\frac{1}{2}\left[\frac{\hbar}{2m\sigma^{2}_{0}}\right]^{2}x^{2}\, ,
\end{align}
we likewise verify that $\sigma(t_{f})=\sigma_{0}$ with similar numerical error. If the overdensity is initially displaced by $\overline{x}_{0}$ from the origin, as expected the center oscillates within the potential well as
\begin{align}
    \overline{x}(t) = \overline{x}_{0} \cos\left[\frac{(t-t_{0})\hbar}{2m\sigma_{0}^{2}}\right]\, .
\end{align}
We confirm the ideal temporal convergence of both numerical schemes by comparing spatially averaged deviations to the analytical solution after one oscillation, shown in \autoref{fig:2vs6}. Note that $\Delta t = 10^{-2}$ corresponds to the maximum time step allowed by stability criteria for a comparable finite difference simulation.

\subsection{Isolated soliton}
\label{appendix:isolatedSoliton}
As the ground state of the Schr\"{o}dinger-Poisson system, a soliton is stationary in time. However, any initial or numerically evolved configuration can only approach the analytical ground state.  \autoref{fig:3} shows that small perturbations result in a periodic ringing of the soliton, which contracts and expands. While we recover the correct quasinormal frequency \cite{Veltmaat2018}
\begin{align}
    \nu = 10.94\left(\frac{\rho_{c}}{10^{9}M_{\odot}\text{kpc}^{-3}}\right)^{1/2}\text{Gyr}^{-1}\, ,
\end{align}
the relative change in potential and kinetic energy depends on the size of the perturbation and thus on resolution. Here, $\rho_{c}$ is the central soliton density.
 
\begin{figure}[b]
\centering
  \includegraphics[width=\linewidth]{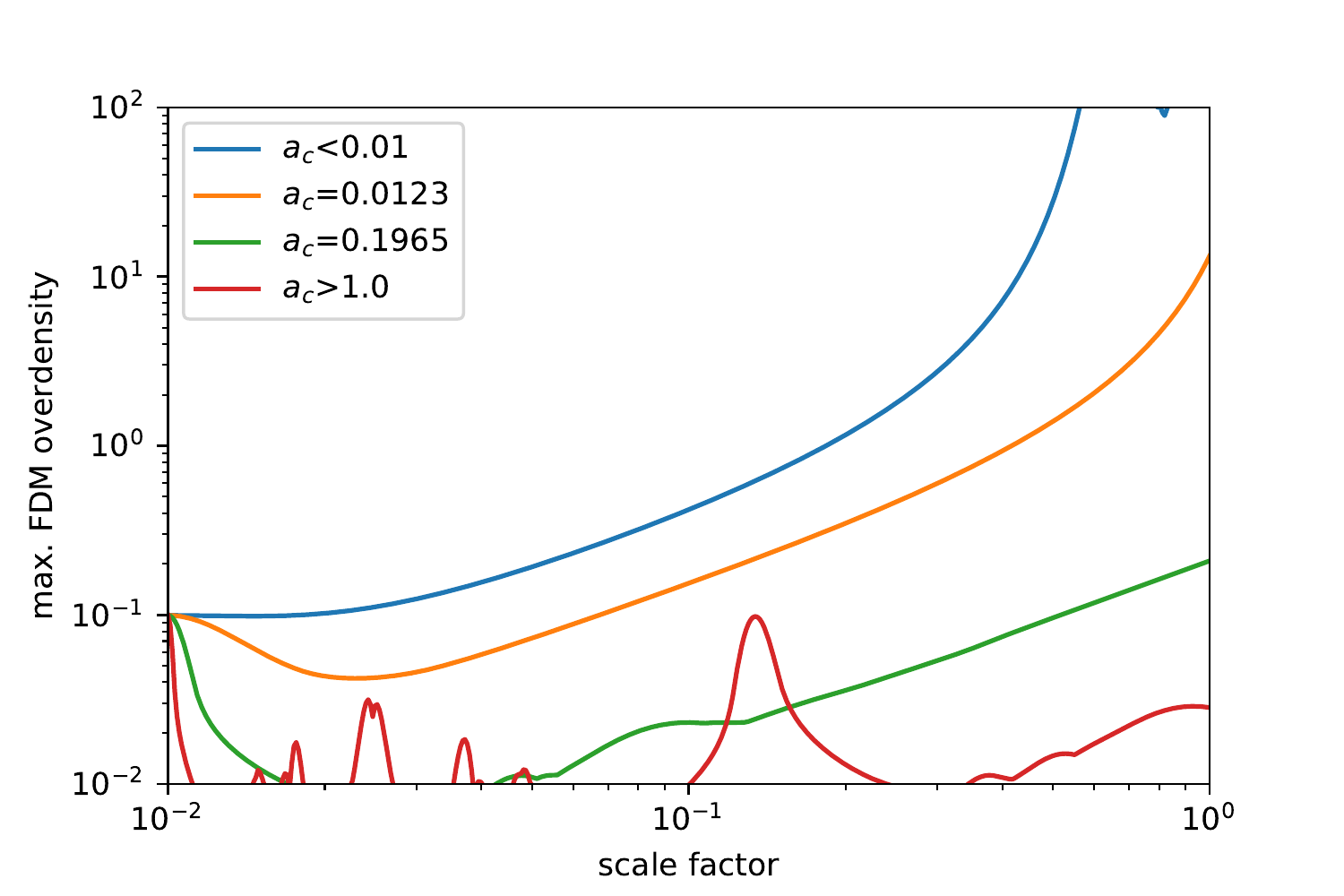}
    \caption{Evolution of the maximum FDM over-density during spherical collapse beginning as soon as the mass inside the simulation box exceeds the Jeans mass at a scale factor $a_{c}$. \label{fig:6}}
\end{figure}   

\subsection{Boson Star Condensation}
\label{appendix:iBScondensation}
The condensation of a Boson star from random Gaussian initial fluctuations, as first investigated in \cite{Levkov2018a}, provides a stringent test of the code. The initial conditions were set up on a $64^3$ grid with a side length of $L=30 \lambda_{\rm dB}$. The overall amplitude of the fluctuations was scaled such that the total mass in the simulation box is $N=10$; we refer the reader to the original paper for details \cite{Levkov2018a}. \autoref{fig:2} shows the expected linear increase in maximum amplitude after one condensation time.

\subsection{Linear mode evolution}
\label{sec:lme}
A one-dimensional, self-gravitating, sinusoidal over-density in a universe containing 100\% FDM obeys equation~\ref{eq:lingr2a} with the RHS set to zero. We observe the expected evolution of different growing modes \cite{Marsh2016a}
\begin{align}
    \label{eq:gm}
\delta_{+}(k,a) = \frac{3}{\tilde{k}^{2}}\sin\left(\tilde{k}^{2}\right)+\left[\frac{3}{\tilde{k}^{4}}-1\right]\cos\left(\tilde{k}^{2}\right)\, ,
\end{align}
where $\tilde{k}=k/\sqrt{\sqrt{a}mH_{0}/\hbar}\propto k/k_{\rm J}(a)$ is inversely proportional to the Jeans scale $k_{\rm J}(a)$. While modes with wave number $k<k_{\rm J}$ grow linearly with scale factor $a$, modes with $k>k_{\rm J}$ oscillate in time. Intermediate modes just below $k_{\rm J}$ oscillate until they cross the Jeans scale after which they start to grow linearly. All three scenarios can be seen in \autoref{fig:5}.

\subsection {Spherical collapse of fuzzy dark matter}
\label{sec:sphcol}

We investigated the spherical collapse of a Gaussian over-density with initial maximum density contrast $\delta=0.1$ above the critical density of a flat, matter dominated universe. For all runs, the over-density was placed centrally in a comoving box containing the $8\sigma$ region of the Gaussian perturbation. We thus scale the over-density by changing the box size. The Jeans mass below which collapse is suppressed can be approximated as \cite{Marsh2016b}
\begin{align}
    M_{J}(a) =& 3.4\times 10^{8}\left(\frac{m}{10^{-22}\text{\,
    eV}}\right)^{-3/2}\left(\frac{\Omega_{m}h^{2}}{0.14}\right)^{1/4}\nonumber\\
    &\times 
    h^{-1}a^{-3/4}M_{\odot}\nonumber\\
    \simeq& 1.7\times 10^{8}a^{-3/4}M_{\odot},
\end{align}
where we used $\Omega_{m}=1$, $h=0.7$, $m = 2.5\times 10^{-22}\text{\, eV}$. From \autoref{fig:6} we see that collapse occurs once the mass inside the box starts to exceed $M_{J}(a)$.

\end{document}